\documentclass{article}
\usepackage{spconf,amsmath,graphicx}

\usepackage{bm}
\usepackage{amsmath}  
\usepackage{amssymb} 
\usepackage{multirow}
\usepackage{makecell}  
\usepackage{booktabs}  
\usepackage{float}  
\usepackage{balance} 
\usepackage{cite}  

\title{REAL-TIME SPEECH ENHANCEMENT WITH DYNAMIC ATTENTION SPAN}
\name{Chengyu\ Zheng$^{\dag*}$\thanks{*Work was done during the internship at MSRA.}, Yuan\ Zhou$^{\ddag}$, Xiulian\ Peng$^{\ddag}$, Yuan\ Zhang$^{\dag}$, Yan\ Lu$^{\ddag}$}
\address{$^{\dag}$Communication University of China, Beijing, China \\
$^{\ddag}$Microsoft Research Asia, Beijing, China \\}
\begin{document}
%
\maketitle
\begin{abstract}
For real-time speech enhancement (SE) including noise suppression, dereverberation and acoustic echo cancellation, the time-variance of the audio signals becomes a severe challenge. The causality and memory usage limit that only the historical information can be used for the system to capture the time-variant characteristics. We propose to adaptively change the receptive field according to the input signal in deep neural network based SE model. 
Specifically, in an encoder-decoder framework, a dynamic attention span mechanism is introduced to all the attention modules for controlling the size of historical content used for processing the current frame. Experimental results verify that this dynamic mechanism can better track time-variant factors and capture speech-related characteristics, benefiting to both interference removing and speech quality retaining.
\end{abstract}
%
%
\section{Introduction}

Acting as a preprocessor in teleconferencing, speech enhancement (SE) including noise suppression (NS), dereverberation and acoustic echo cancellation (AEC), aims at removing interference, such as additive noise, reverberation and acoustic echo.
Real-time SE requires the causality, limited memory usage and low computational complexity. 
Causality means the system can only use the current and historical audio frames when processing the current frame. Memory usage restricts the model size and the historical content used. Computational complexity is the runtime cost when running the SE system.

The time-variance of the audio signals recorded by the microphone becomes a severe challenge for the real-time SE, since the causality and memory usage constrains the system to capture the time-variant characteristics only from the limited historical information. The challenging cases frequently happen in the varying time delay between microphone and reference signals and the changing acoustic environments. 
To handle this challenge, the traditional SE methods utilize adaptive filters to trace the time-variant factors \cite{haykin2014adaptive, kao2003design} and some recent methods propose to use deep neural networks (DNN) for leveraging their powerful nonlinear modeling capacity, either cascading them as a post-filter after the adaptive filters for further removing the residual echo and noise \cite{shu2021joint,valin2021low,peng2021acoustic,sun2022explore,zhang2022multi}, or using them in an end-to-end way to directly model the mapping between the microphone signal and the target clean speech \cite{westhausen2021acoustic,zheng21ft_interspeech,watcharasupat2022end,yu2022neuralecho,cui2022multi}. 
However, the size of historical content used for processing each input frame are fixed due to the weights and receptive field of the DNN are usually frozen during inference, which makes the model can not explicitly capture the time-variant characteristics including both the environmental interference and speech-related features.

In this paper, we propose to explicitly capture the time-variance by generating the input-adaptive receptive field for the DNN-based SE model. The SE model consists of two individual encoders, one temporal attention (TA) merge module, one decoder and several repeated modules inserted in-between them (Fig. \ref{fig_overall}). 
The repeated module contains a temporal convolution module (TCM) and a group-wise self-attention (GTSA) module.
A dynamic attention span (DAS) is introduced to all the attention modules to adaptively change the receptive field according to each input frame. To save the memory usage, we constrain the DAS within a fixed size instead of all the historical frames. Experimental results show that DAS leads to better performance on time-variant scenarios and on both NS and AEC tasks but with less computational complexity, comparing with the fixed attention span models and other benchmarks.  

\section{Proposed Methods}\label{sec2}

\subsection{Problem Formulation}
\vspace{-0.2cm}

The near-end speech signal $s(t)$ is recorded by the microphone via the acoustic path $h_1(t)$. The reference signal $x(t)$ played by the near-end loudspeaker gets nonlinear distortions $f_{\rm{NL}}$, for example loudspeaker and possible processing distortions, and is also recorded via the acoustic echo path $h_2(t)$. Thus, the time-variant microphone signal $y(t)$ is modeled as: 
\begin{equation}
    y(t)=h_1(t)*s(t)+h_2(t)*f_{\rm{NL}}\left[x(t-\Delta_t)\right]+n(t),
    \label{equa1}
\end{equation}
where $*$ denotes convolution, $\Delta_t$ is the time-variant delay between the reference and microphone signals, 
$h_1(t)$ and $h_2(t)$ are the room impulse response (RIR) which can be represented as the summation of impulse responses for the direct sound, early and late reflection, and $n(t)$ is the additive noise. In this paper, the NS task aims at removing $n(t)$, the AEC task aims at removing the echo generated from $x(t)$, and the dereverberation task aims at removing the early and late reflection generated by $h_1(t)$. 

\begin{figure}
\centerline{\includegraphics[width=0.7\columnwidth]{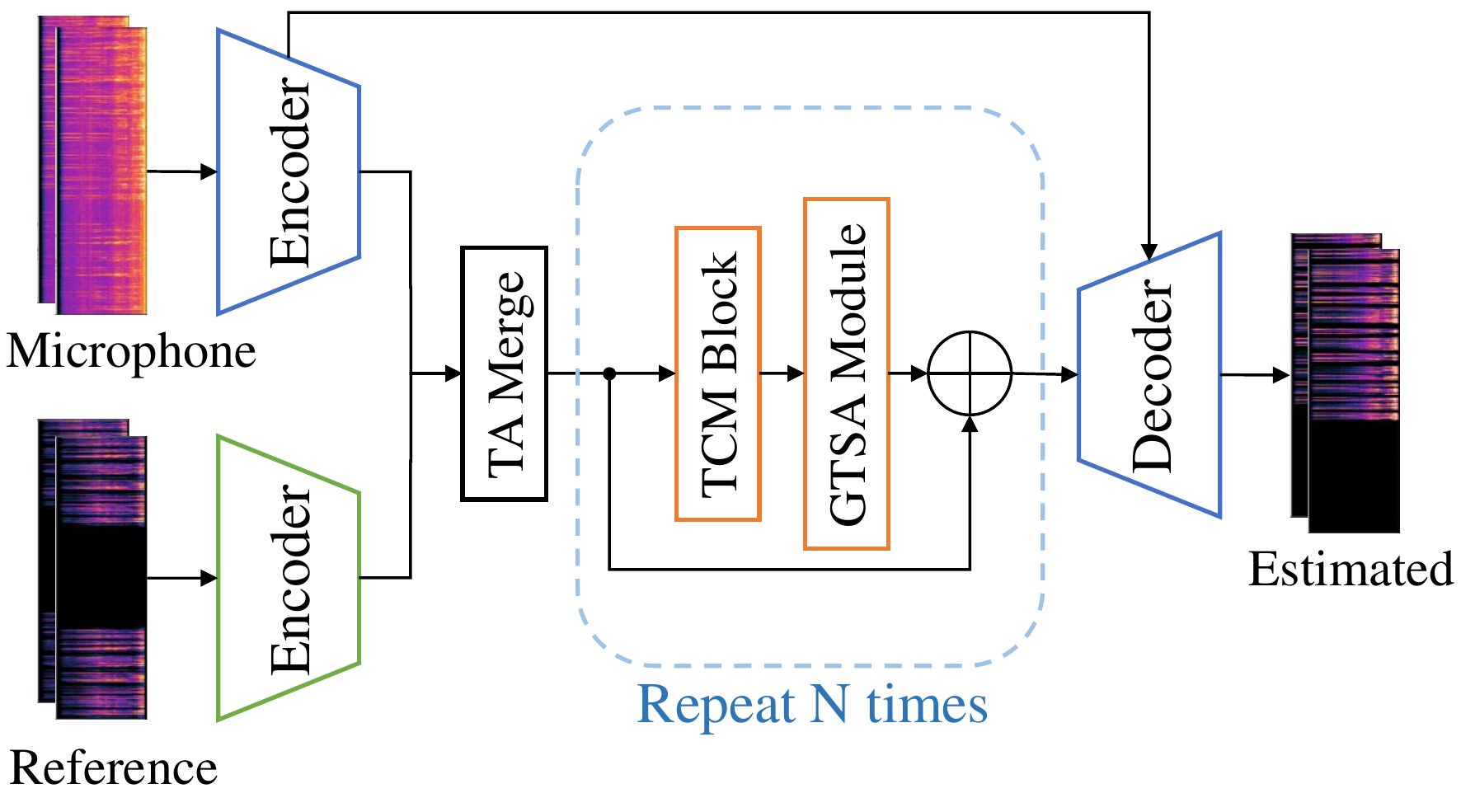}}
\vspace{-.4cm}
\caption{\small{Overall architecture. $N$ is set to 4.}}
\label{fig_overall}
\end{figure}

\vspace{-0.3cm}
\subsection{Model Architecture}
\vspace{-0.2cm}

Fig. \ref{fig_overall} shows the overall architecture of the proposed model, which consists of two individual encoders, one TA merge module, one decoder and several repeated modules connecting between them. The model takes the Short-time Fourier Transform (STFT) spectrums of both the microphone $\mathcal{Y}\in \mathbb{R}^{2\times T\times F}$ and the reference $\mathcal{X}\in \mathbb{R}^{2\times T\times F}$ signals as the inputs and estimates the STFT spectrum of the near-end speech signal, where $T$ is the number of time frames, $F$ is the number of frequency bins. Each complex spectrum has real and imaginary parts.

\textbf{Encoders and Decoder.} The microphone and reference spectrums are input to two individual encoders, respectively. The structure of the encoders and decoder are the same as that in \cite{zheng2022time}. 



\textbf{TA Merge Module.} Due to the time-misalignment between reference and microphone signals, temporal attention (TA) across the two signals is introduced to explicitly capture the cross-correlations between them and merge the features from dual pathways into one. Specifically, taken the microphone feature $\mathcal{F}^{\rm{Mic}}\in\mathbb{R}^{C\times T\times F'}$ and the reference feature $\mathcal{F}^{\rm{Ref}}\in\mathbb{R}^{C\times T\times F'}$ from two corresponding encoders, the TA is given by: 
\begin{equation}
    \begin{aligned}
        &\mathcal{F}^{Q} = {\rm Reshape}({\rm Conv2D}^{Q}(\mathcal{F}^{{\rm Mic}})), \\
        &\mathcal{F}^{k} = {\rm Reshape}({\rm Conv2D}^{k}(\mathcal{F}^{{\rm Ref}})),k\in\{K,V\}, \\
        &\mathcal{F}^{TA} = {\rm Softmax}({\rm Mask}(\mathcal{F}^{Q}\cdot(\mathcal{F}^{K})^{\rm tr}/\sqrt{F'C}))\cdot \mathcal{F}^{V}, 
    \end{aligned}
    \label{equa2}
\end{equation}
where $\mathcal{F}^{k}\in\mathbb{R}^{T\times C'}$ ($k\in\{Q,K,V\}$), and $\mathcal{F}^{TA}\in\mathbb{R}^{T\times C'}$, respectively. $\rm{Reshape}$ denotes a tensor reshape from $\mathbb{R}^{C\times T\times F'}$ to $\mathbb{R}^{T\times C'}$ ($C'=F'C$). ${\rm Conv2D}$ represents a 2-D convolutional layer with kernel size of (1,1), stride of (1,1) and channel number of $C$, followed by a batch normalization \cite{ioffe2015batch} and a parametric ReLU \cite{he2015delving}. The superscription $\rm{tr}$ means transposing the last two dimensions of the tensor. $\rm{Mask}$ denotes the masking operation \cite{zheng2022time} with an attention span of $T_w$ to keep causality and fix the largest attention span to $T_w$.

\textbf{Repeated Module.} The repeated module contains a TCM defined in \cite{pandey2019tcnn} and a GTSA module \cite{zheng2022time}, as depicted in the dotted box in Fig. \ref{fig_overall}. The incorporation of the TCM and the GTSA module aims at capturing long/short-term dependencies along the temporal dimension in parallel. 

\vspace{-0.3cm}
\subsection{DAS Mechanism}
\vspace{-0.2cm}

DAS mechanism is introduced to all the attention modules (TA merge and GTSA) to capture the time-variant factors by adjusting input-adaptive receptive field. Inspired from \cite{sukhbaatar2019adaptive}, for input feature $\bm{x}_t$, we compute a DAS $z_t$ indicating the historical frames used for attention. This can be represented as: 
\begin{equation}
    \begin{aligned}
        &z_t = T_w\sigma(\bm{v}^{\rm tr}\bm{x}_t+b), \\
        &m_{z_t}(t,r) = {\rm min}\left[{\rm max}\left[\frac{1}{R}\left(R+z_t-(t-r)\right),0\right],1\right], \\
        &a_{t,r} = \frac{m_z(t,r){\rm exp}(s_{t,r})}{\sum_{r=t-T_w+1}^{t}m_z(t,r){\rm exp}(s_{t,r})},
    \end{aligned}
\end{equation}
where the vector $\bm{v}$ and the scalar $b$ are learnable parameters. In the TA merge module, $\bm{x}_t\in\mathbb{R}^{2C'\times 1}$ is obtained by concatenating the reference and microphone features along the channel dimension, and $\bm{v}\in\mathbb{R}^{2C'\times 1}$. In the GTSA module, $\bm{x}_t\in\mathbb{R}^{C'\times 1}$ is the output feature from the previous repeated module, and $\bm{v}\in\mathbb{R}^{C'\times 1}$. $\sigma$ denotes the Sigmoid function mapping the value to the range of $(0,1)$ to constrain $z_t\in(0,T_w)$, so that the DAS is within a fixed size instead of using all the historical frames. To keep causality and ensure the gradient during back propagation, $\rm{Mask}$ in Equation (\ref{equa2}) is defined as a soft masking function $m_z(t,r),t\geq r$ for attention with DAS. $R$ is a hyper-parameter that controls the softness. $s_{t,r}$ denotes the similarity score between the query vector at $t$-th frame and the key vector at $r$-th frame, and $a_{t,r}$ denotes the attention score after Softmax operation. 

\section{Experiment Settings}

\subsection{Datasets}
\vspace{-0.2cm}

We synthesize 1166.7 hours of audio samples for training and 9.7 hours for validation, using speech and noise from the Interspeech 2021 DNS Challenge and RIRs from the Interspeech 2021 AEC Challenge, as that mentioned in \cite{zheng2022time}. The ratio of data for far-end single-talk (FST), near-end single-talk (NST) and double-talk (DT) is 1:1:5. 

For the AEC task, a synthetic test set \cite{zheng2022time} for ablation study and the real-recorded blind test set of AEC Challenge ICASSP 2022 \cite{cutler2022AEC} are used. For the NS task, 738 clips with the tag ``Primary'' from the blind test set of Track-1 non-personalized DNS at DNS Challenge ICASSP 2022 \cite{dubey2022icassp} are used. 
All real-recorded clips are originally collected at a sampling rate of 48 kHz and resampled to 16 kHz. 


\begin{table}
\centering
\caption{\small{Ablation study on AEC with time-variant factors.}}
\small
\resizebox{\columnwidth}{!}{
    \begin{tabular}{lcccccc}
        \hline
        \multirow{2}{*}{Model} & \multicolumn{2}{c}{DAS in}& \multirow{2}{*}{ERLE} & \multirow{2}{*}{PESQ} & \multirow{2}{*}{Para.(M)} & \multirow{2}{*}{MACs(M/sec)} \\ 
        \cline{2-3}
         & TA & GTSA & & & & \\
        \hline
        Unprocessed & - & - & -& $1.483$ & -& - \\
        Baseline-Cat & $\times$ & $\times$ & $39.220\pm 1.976$& $2.244\pm 0.033$ & 1.970 & 462.10\\
        Baseline-TA & $\times$ & $\times$ & $42.263\pm 2.036$& $2.182\pm 0.057$ & 1.974 & 465.64  \\
        TA-DAS & $\checkmark$ & $\times$ & $43.946\pm 0.963$& $2.196\pm 0.049$ & 1.975 & $<$465.64 \\
        GTSA-DAS & $\times$ & $\checkmark$ & $42.181\pm 1.461$& $\textbf{2.289}\pm \textbf{0.019}$ & 1.975 & $<$465.64 \\
        All-DAS & $\checkmark$ & $\checkmark$ & $\textbf{44.732}\pm \textbf{1.183}$& $2.223\pm 0.051$ & 1.976 & $<$465.64 \\
        \hline
    \end{tabular}
}
\label{table_ablation_aec_synth}
\end{table}

\begin{table}
\centering
\vspace{-0.6cm}
\caption{\small{Ablation study on NS.}}
\small
\resizebox{0.7\columnwidth}{!}{
    \begin{tabular}{lccc}
        \hline
        Model & SIG & BAK & OVRL \\ 
        \hline
        Unprocessed & ${3.972}$ & $2.043$ & $2.344$ \\
        Baseline-Cat & $3.228\pm 0.012$ & $4.026\pm 0.011$ & $2.955\pm 0.011$ \\
        Baseline-TA & $3.223\pm 0.013$ & $4.037\pm 0.005$ & $2.954\pm 0.011$ \\
        TA-DAS & $3.238\pm 0.012$ & $4.041\pm 0.009$ & $2.971\pm 0.012$ \\
        GTSA-DAS & $3.247\pm 0.014$ & $4.046\pm 0.012$ & $2.981\pm 0.013$ \\
        All-DAS & $\textbf{3.259}\pm \textbf{0.011}$& $\textbf{4.048}\pm \textbf{0.009}$ & $\textbf{2.993}\pm \textbf{0.013}$ \\
        \hline
    \end{tabular}
}
\label{table_ablation_dns}
\end{table}

\vspace{-0.4cm}
\subsection{Implementation Details and Baselines}
\vspace{-0.2cm}

The implementation details are similar to \cite{zheng2022time}. 
All signals are transformed to time-frequency domain using a 20-ms Hanning window, 10-ms overlap and 320-point DFT. 
The fixed-size attention span $T_{w}$ is set to 100, and the hyper-parameter $R$ is set to 2. All models are trained for 100 epochs with a batch size of 200. 
For the NS task, the input signal of the reference encoder is set to zero, so that the model can adapt for both NS and AEC tasks. Ablation experiments are conducted on the following five configurations: 

(1) \textbf{Baseline-Cat}: baseline model mentioned in \cite{zheng2022time}.

(2) \textbf{Baseline-TA}: proposed model with a fixed-size attention span $T_w$ in both TA merge and GTSA modules. 

(3) \textbf{TA-DAS}: proposed model with DAS only in the TA merge module.

(4) \textbf{GTSA-DAS}: proposed model with DAS only in all GTSA modules.

(5) \textbf{All-DAS}: proposed model with DAS in both TA merge module and all GTSA modules.


\vspace{-0.4cm}
\subsection{Evaluation Metrics}
\vspace{-0.2cm}

For the synthetic test set, the echo return loss enhancement (ERLE) \cite{enzner2014acoustic} is used for FST periods and the perceptual evaluation of speech quality (PESQ) \cite{rec2005p} is used for DT periods. 

For the real-recorded test sets, we use the AECMOS tool \cite{purin2022aecmos} with regards to echo ratings (ECHO) and other degradation ratings (OTHER), and DNSMOS tool \cite{reddy2021dnsmos} with regards to speech degradation (SIG), noise suppression (BAK) and the overall quality (OVRL) to evaluate all the methods.

\section{Experimental Results}

\subsection{Ablation Study}
\vspace{-0.2cm}

\begin{figure}[t]
\centerline{\includegraphics[width=0.9\columnwidth]{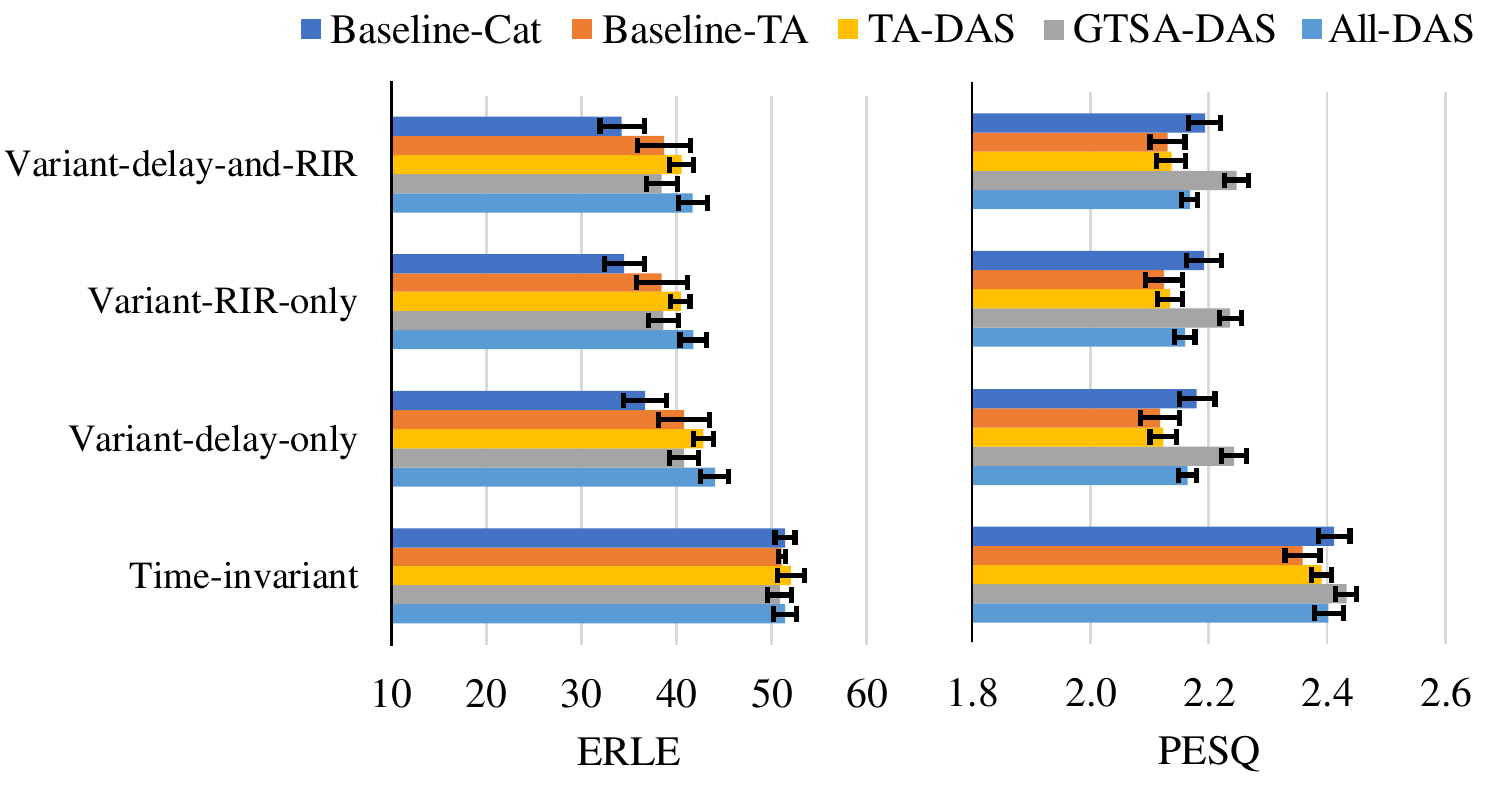}}
\vspace{-.4cm}
\caption{\small{ERLE and PESQ results of different scenarios on the synthetic test set. For the time-invariant, variant-delay-only, variant-RIR-only and variant-delay-and-RIR scenarios, the PESQ values of the unprocessed signals are 1.548, 1.455, 1.462, 1.464, respectively.}}
\label{fig_ablation_ERLE_PESQ}
\end{figure}

Table \ref{table_ablation_aec_synth} shows the results on the AEC synthetic test set. Fig. \ref{fig_ablation_ERLE_PESQ} shows the evaluation metrics under time-invariant/variant scenarios. 
We have following conclusions: (1) Baseline-TA significantly outperforms Baseline-Cat on ERLE metric in all time-variant cases, indicating that TA merge module helps the model to better remove the time-variant echo. However, the PESQ metric of Baseline-TA decreases in both time-invariant/variant cases, indicating that only introducing TA merge module is not enough for balancing the echo cancellation and speech quality retaining. 
(2) On top of Baseline-TA, introducing DAS to the TA merge module slightly brings gains on both ERLE and PESQ. 
(3) The GTSA-DAS model keeps the stable improvement of Baseline-TA on ERLE and shows much better PESQ. This indicates that introducing DAS to the GTSA module enhances the model on capturing more speech-related time-variant characteristics, which leads to better near-end speech quality in DT scenario.
(4) All-DAS model obtains the largest ERLE in all these cases among all the models, while the PESQ degrades compared with the GTSA-DAS model. These show that introducing DAS to TA and GTSA simultaneously can significantly improves the echo cancellation but with this simple cascading of DAS-enabled TA and GTSA might not add up to their respective advantages on speech quality retaining in AEC task.

\begin{figure}[htb]
\begin{minipage}[b]{\linewidth}
  \centering
  \centerline{\includegraphics[width=0.65\linewidth]{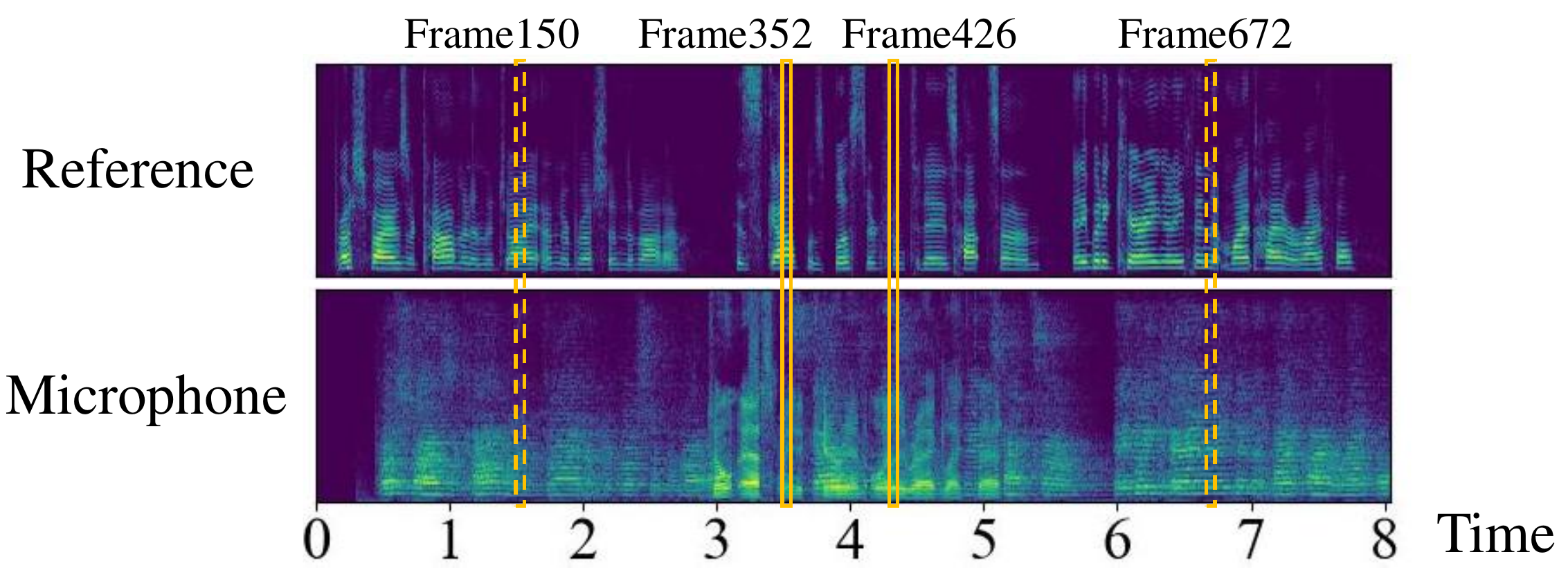}}
  \center{\small{(a) From top to bottom are log-power spectrograms of the reference and microphone signals, respectively. }}\medskip
  \vspace{-0.1cm}
\end{minipage}
\hfill
\begin{minipage}[b]{\linewidth}
  \centering
  \centerline{\includegraphics[width=\linewidth]{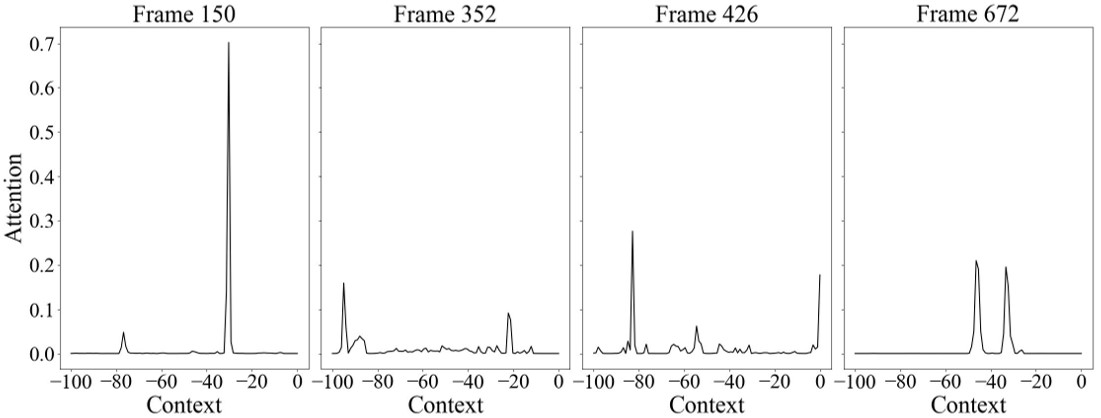}}
  \center{\small{(b) Attention scores of the TA merge module in Baseline-TA.}}\medskip
\end{minipage}
\hfill
\begin{minipage}[b]{\linewidth}
  \centering
  \centerline{\includegraphics[width=\linewidth]{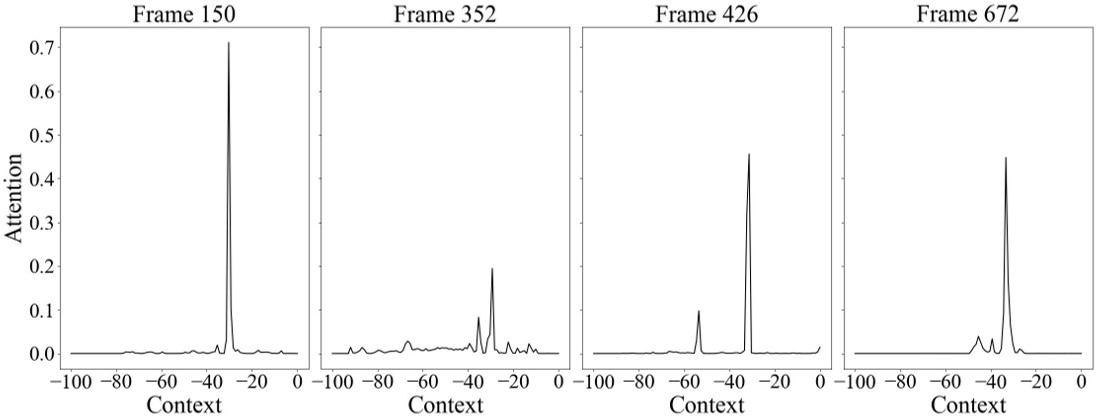}}
  \center{\small{(c) Attention scores of the TA merge module in All-DAS model. From left to right, the attention spans are 74.59, 89.90, 84.40 and 68.27, respectively.}}\medskip
  \vspace{-.5cm}
\end{minipage}
\caption{\small{Visualization of the TA merge module attention patterns.} 
}
\label{fig_attn_pat_casewise}
\end{figure}

Table \ref{table_ablation_dns} shows the results on the NS blind test set. Different from AEC, introducing DAS to all the attention modules (All-DAS) simultaneously shows the best result, which verifies that the DAS mechanism can bring stable benefit for removing additive environmental noise.

The total algorithmic latency of all the models is 30 ms. On top of Baseline-TA, introducing DAS slightly increases the model parameters, but decreases the computational complexity (Table \ref{table_ablation_aec_synth}).

\vspace{-0.3cm}
\subsection{Impact of DAS}
\vspace{-0.2cm}

To explore how the DAS brings benefit to the model, we synthesize an input-pair of the reference and microphone signals with 300-ms delay (i.e. 30 frames in the spectrum) and visualize the attention patterns of TA merge module for 4 picked frames. As shown in Fig. \ref{fig_attn_pat_casewise} (a), frame 150 and frame 672 are from FST (highlighted in dotted boxes), and frame 352 and frame 426 are from DT (highlighted in solid boxes). Fig. \ref{fig_attn_pat_casewise} (b) and (c) show the corresponded attention scores of the TA merge module in the Baseline-TA and All-DAS models, respectively. 
The ERLE and PESQ metrics are 63.34 and 2.35 for the Baseline-TA model, and 63.95 and 2.71 for the All-DAS model. 
The value with a negative symbol in the X-axis denotes the temporal distance between the historical and current frames. The Y-axis shows the attention score. 


For the FST scenario like frame 150, the maximum attention score appears at the $-30^{th}$ frame (equals to the time delay between the two signals) in both the Baseline-TA and All-DAS models, indicating that TA attends to the most related frame from the reference signal for the current microphone frame. However, in frame 672, TA with fixed attention span has two significant peaks in the attention score, while TA with DAS still mainly attends to the $-30^{th}$ frame. 
For the more challenging DT scenario (frames 352 and 426), TA with DAS can still accurately attend to the frame at 300 ms before while Baseline-TA gets confusing attention distributions. These verify that by adjusting the receptive field, DAS improves the accuracy and stability of attention module on capturing the cross-correlations between two signals. 

Table \ref{table_tsa_span} shows the attention span in different channel groups of the last GTSA module in the All-DAS model. With the structure of channel grouping, the DAS gives the model different temporal dependencies in different channel groups. For example, for frame 352, the G1 and G3 give short attention span, while the G2, G4 and G5 give relatively long. This observation indicates that the GTSA-DAS can adjust the receptive field at different channels automatically, which shows the potential of the module to be further designed as a foundational component of the DNNs on audio processing for its adaptive short/long-term feature capturing ability. 


\begin{table}
\centering
\caption{\small{Attention span in different groups of the last GTSA module in the All-DAS model.}}
\small
\resizebox{0.65\columnwidth}{!}{
    \begin{tabular}{cccccc}
        \hline
        Group & G1 & G2 & G3 & G4 & G5 \\ 
        \hline
        Frame 150 & 0.00 & 100.00 & 94.88 & 99.94 & 100.00 \\
        Frame 352 & 27.93 & 100.00 & 32.01 & 99.98 & 99.98 \\
        Frame 426 & 92.94 & 100.00 & 23.08 & 100.00 & 99.98 \\
        Frame 672 & 99.43 & 100.00 & 55.64 & 99.76 & 99.99 \\
        \hline
    \end{tabular}
}
\label{table_tsa_span}
\end{table}

\begin{table}
\centering
\vspace{-0.6cm}
\caption{\small{Comparison on the AEC real-recorded blind test set.}}
\small
\resizebox{0.9\columnwidth}{!}{
    \begin{tabular}{ccccccc}
        \hline
        \multirow{2}{*}{Methods}& \multicolumn{2}{c}{FST}& \multicolumn{2}{c}{DT}& NST& \multirow{2}{*}{Para.(M)} \\ 
        \cmidrule(lr){2-3} \cmidrule(lr){4-5} \cmidrule(lr){6-6}
         & ERLE& ECHO& ECHO& OTHER& OTHER&  \\
        \hline
        Unprocessed& -& 2.046& 1.812& 4.110& 3.940& - \\
        SpeexDSP& 11.319& 2.846& 2.853& \textbf{4.154}& 4.081& - \\
        NSNet& 55.059& 4.519& 4.235& 3.558& \textbf{4.110}& 1.30 \\
        DTLN-AEC (S) & 32.691& 4.204& 3.917& 3.499& 3.916& 1.8 \\
        DTLN-AEC (M) & 33.004& 4.292& 4.046& 3.601& 4.012& 3.9 \\
        DTLN-AEC (L) & 37.918& 4.351& 4.236& 3.687& 4.026& 10.4 \\
        All-DAS & \textbf{60.107}& \textbf{4.662}& \textbf{4.620}& 3.912& 3.943& 1.976 \\
        \hline
    \end{tabular}
}
\label{table_comparison}
\end{table}

\vspace{-0.3cm}
\subsection{Comparison with Other Methods}
\vspace{-0.2cm}

Table \ref{table_comparison} shows the results on the ICASSP 2022 AEC challenge blind test set of different methods.  
SpeexDSP\footnote{https://github.com/xiongyihui/speexdsp-python} is a non-DNN-based method. 
NSNet\footnote{https://github.com/microsoft/AEC-Challenge} is the baseline model at AEC Challenge ICASSP 2022, and DTLN-AEC\footnote{https://github.com/breizhn/DTLN-aec} \cite{westhausen2021acoustic} is one of the top-5 models at AEC Challenge ICASSP 2021.

From the results, we find that compared to All-DAS model: (1) SpeexDSP fails to cancel the echo for its limited nonlinear modeling ability. (2) NSNet outperforms most methods in NST but losses the effectiveness in DT, indicating that NSNet fails to accurately capture the target speech-related characteristics to keep the balance between echo cancelling and speech retaining. 
(3) DTLN-AEC tends to preserve both the far-end and near-end signals with strong echo leaking. 


\section{Conclusions}

In this paper, we propose to introduce DAS mechanism into the attention module for DNN-based real-time SE. 
Experimental results show that model with DAS can better track time-variant factors and capture speech-related characteristics from the input audio, benefiting to both interference removing and speech quality retaining. 
Also, the DAS improves the TA merge module to accurately capture time-variant correlations between the reference and microphone signals, especially for better robustness on the AEC task. The input-adaptive receptive field enables the GTSA module to dynamically capture the long/short-term dependencies in parallel. 

\bibliographystyle{IEEEbib}
\small
\bibliography{refs}

\end{document}